\documentstyle[prl,aps,preprint]{revtex}
\newcommand{\eq}{\begin{equation}}
\newcommand{\ee}{\end{equation}}
\newcommand{\eqa}{\begin{eqnarray}}
\newcommand{\eea}{\end{eqnarray}}
\def\wpsi{{\bar{\psi}}}
\def\tp{{t_\perp}}
\def\tpc{{t_\perp^c}}
\def\sm{{NL$\sigma$M}}

\def\sxxtd{{\sigma_{xx}^{2d}}}
\def\sxytd{{\sigma_{xy}^{2d}}}
\def\sxx{{\sigma_{xx}}}
\def\sxxc{{\sigma_{xx}^c}}

\def\sxy{{\sigma_{xy}}}
\def\sxxp{{\sigma_{xx}^\prime}}
\def\sxyp{{\sigma_{xy}^\prime}}
\def\suvp{{\sigma_{\mu\nu}^\prime}}
\def\sxxo{{\sigma_{xx}^0}}

\def\szzo{{\sigma_{zz}^0}}
\def\sxyo{{\sigma_{xy}^0}}

\def\uv{{\epsilon_{\mu\nu}}}
\def\dt{{\partial_\tau}}
\def\dx{{\partial_x}}
\def\dmu{{\partial_\mu}}
\def\dnu{{\partial_\nu}}
\def\dplus{{\partial_+}}
\def\dminus{{\partial_-}}
\def\drho{{\partial_\rho}}
\def\tr{{\rm tr}}
\def\Tr{{\rm Tr}}
\def\dtheta{{\Delta\theta}}
\def\lp{{\lambda^\prime}}
\begin{document}
\draft
\title{
Localization and Metal-Insulator Transition 
in Multi-Layer Quantum Hall Structures}
\author{Ziqiang Wang}
\address{Department of Physics, Boston College, Chestnut Hill, MA 02167}
\maketitle
\begin{abstract}
We study the phase structure and Hall conductance
quantization in weakly coupled multi-layer electron systems in the integer 
quantum Hall regime. We derive an effective field theory
and perform a two-loop renormalization group calculation.
It is shown that (i) finite interlayer tunnelings (however small)
give rise to successive metallic and insulating phases 
and metal-insulator transitions in the unitary universality class.
(ii) The Hall conductivity is not renormalized in the metallic phases
in the 3D regime. (iii) The Hall conductances are quantized in 
the insulating phases. In the bulk quantum Hall phases, the 
effective field theory describes the transport on the surface.
\end{abstract}
\newpage
The quantum Hall effect (QHE) in a two-dimensional electron gas (2DEG)
has led to many new physical concepts and principles \cite{qhe}. 
The main part of the phenomenology is that in strong magnetic fields,
a 2DEG exhibits continuous zero-temperature phase transitions 
between successive quantum Hall states of vanishing dissipation and
quantized Hall resistances. It is interesting to ask what 
happens to the physics associated with the QHE in dimensions greater than two
\cite{bh}. Experimentally, two classes of quasi-three-dimensional
materials have been found to show integer-quantized Hall plateaus:
multi-layer quantum wells or superlattices formed
by GaAs/AlGaAs graded heterostructures \cite{stormer,brookswang} and 
molecular crystals (TMTSF)$_2$X (X=PF$_6$,ClO$_4$) 
\cite{organics}. In this paper, we concentrate on the former which is
a natural generalization of the QHE above two dimensions. 
Thus, the changes in the phase structure and the properties of
the phase transitions in quantum Hall layers coupled by weak
interlayer tunnelings are the concerns of the present paper.
Moreover, we focus on the integer quantum Hall regime and ignore
the effects of electron-electron interactions \cite{leewanginter}.

We shall follow the approach of Chalker and Dohmen who generalized
the Chalker-Coddington network model \cite{cc} for the integer QHE (IQHE)
in a 2DEG to layers of networks coupled by interlayer tunnelings \cite{cd}.
The advantage of this approach is that the single layer network model is known
to correctly describe the universality class of the 2D integer quantum Hall 
transitions. Chalker and Dohmen performed 
numerical transfer matrix calculations and demonstrated the existence of 
three phases: insulator, metal, and quantized Hall conductor, and extended 
surface states in the quantized Hall state. In this paper we provide 
an analytical treatment using the effective field theory representation
of the network model \cite{lee,leewanghubbard,km}. 
We first show that the long wavelength transport properties are governed
by a 3D anisotropic unitary nonlinear $\sigma$-model (\sm). The anisotropic
couplings are the dissipative conductivities, whereas the
Hall conductivity enters as a coupling to the layered sum of the 2D 
topological term. The renormalization group (RG) is then
used to study the crossover between two and three dimensions.
In the 3D regime, the RG flow equations for the conductances are calculated
to two-loop order to determine the phase structure in the
plane spanned by the Fermi energy and the interlayer tunneling.
The results show that a finite interlayer coupling (however weak) leads to
metallic and insulating phases and metal-insulator transitions  
in the unitary universality class. Furthermore, the Hall conductivity is found
to be unrenormalized by localization effects in the 3D regime. 
We show that the Hall conductance is quantized in the insulating phases
provided that the above results hold to all orders in the RG.
Finally, we demonstrate that in the
quantum Hall phases, the field theory reduces to
the one appropriate for the coupled edge states on the surface.

Following Ref.~\cite{leewanghubbard},
the Hamiltonian for $N$ layered networks in the $(x,y)$-plane coupled
in the $z$-direction by interlayer tunneling $\tp$ is,
\eqa
H_0=&&\sum_{x,z}(-1)^x\!\!\int\!dy
\psi^\dagger(x,y,z)\left[i{\partial_y}-V_{x,y,z}\right]\psi(x,y,z) 
\nonumber \\
-&&\sum_{x,z}\int\!dy \ t_x
\left[\psi^\dagger (x+1,y,z)\psi(x,y,z) + {\rm h.c.}\right]
\nonumber \\
-&& \sum_{x,z}\int\!dy \ \tp\left[\psi^\dagger(x,y,z+1)\psi(x,y,z) + {\rm h.c.}\right].
\label{h0}
\eea
Here $\psi^\dagger$ creates an electron traversing the edges of
the Hall droplets as modeled by the links of the network. 
$t_x=t[1-\delta(-1)^x]$, where $\delta$ measures the distance of the
Fermi energy ($E_F$) relative to the center of the Landau level ($E_c$),
represents the quantum tunneling amplitudes at the saddle points of 
the random potential, {\it i.e.} at the nodes of the network.
$V$ is a local random variable that generates the link
Aharonov-Bohm phases, $<V_{x,y,z}V_{x^\prime,y^\prime,z^\prime}>=
U\delta_{x,x^\prime}\delta_{z,z^\prime}\delta(y-y^\prime)$.

For $\tp=0$, Eq.~(\ref{h0}) describes $N$ decoupled 2D networks, each
undergoes a quantum Hall transition as $\delta$ is varied.
In this case, quench averaging over $V$ 
and regarding $y$ as the Euclidean time $\tau$,
it has been shown that the original network 
model corresponds to a half-filled 1D U(2n) Hubbard model in 
the limit $n\to0$ \cite{leewanghubbard}.
The 2D quantum Hall transition is then
equivalent to the dimerization transition of the Hubbard chain 
\cite{leewanghubbard}. Generalizing to $\tp\ne0$, we
obtain the generating functional $Z=\int\!{\cal D}[\wpsi,\psi] 
\exp[\int\!d\tau\sum_{xz}(i\eta S_p\wpsi\psi-H_0(\wpsi,\psi))]$ 
in the form of a 2+1-dimensional Euclidean action
if we let $\psi_p\rightarrow\psi_p(i\psi_p)$ and $\wpsi_p\rightarrow
-i \wpsi_p(\bar{\psi}_p)$ for even (odd) $x$ \cite{leewanghubbard},
\eq
S=\int\! d\tau\left[\sum_{x,z,a}\wpsi_a
 \partial_\tau\psi_a+H\vert_{\psi^\dagger(\psi)
 \rightarrow\wpsi(\psi)}\right].
\label{s}
\ee
Here $\eta$ is a positive infinitesimal, $a=(\alpha,p)$ are the replica index 
$\alpha=1,\dots,n$ and energy index $p=+(-)$ for the advanced (retarded) 
channels, and $S_p\equiv{\rm sgn(p)}$. 
The resulting Hamiltonian $H$ in Eq.~(\ref{s}) corresponds to
an interacting quantum theory in two spatial dimensions,
\eqa
H=&-&\sum_{x,z}t_x\left[\psi^\dagger_a(x+1,z)\psi_a(x,z)+{\rm h.c.}\right]
+{U\over 2}\sum_{x,z}\left[\psi^\dagger_a(x,z) 
\psi_a(x,z)\right]^2
\nonumber \\
&+&\sum_{x,z}i\tp(-1)^x\left[\psi^\dagger_a(x,z+1)\psi_a(x,z)+
{\rm h.c.}\right]
-\eta\sum_{x,z}(-1)^xS_p\psi^\dagger_a\psi_a,
\label{h}
\eea
where sums over repeated indices are implied.
Note that Eq.~(\ref{h}) is not the usual quasi-1D U(2n) Hubbard model
for the form of the interchain couplings.

We now derive the effective low energy, long wavelength theory.
The partition function for Eq.~(\ref{h}) can be written as
$Z=\int\!{\cal D}[\wpsi,\psi]{\cal D}[\Delta]
\exp(-{S})$,
\eqa
S&=&\int\! d\tau \sum_{x,z} -\eta (-1)^x S_p \wpsi_a\psi_a
+S_0+S_I+S_\perp
\nonumber \\
S_0&=&\int\!d\tau\sum_{x,z}\left[\wpsi_a\dt\psi_a
-t_x\left(\wpsi_a(x+1)\psi_a(x)+{\rm c.c.}\right)\right],
\nonumber \\
S_I&=&\int\!d\tau\sum_{x,z}\left[{\Delta^2
\over2U}-\left(\wpsi_a\Delta_{ab}\psi_b-{1\over2}\Delta_{ab}
\delta_{ab}\right)\right],
\nonumber \\
S_\perp&=&\int\!d\tau\sum_{x,z}i\tp(-1)^x\left[\wpsi_a(z+1)\psi_a(z)+ 
{\rm c.c.}\right],
\label{s0}
\eea
where $\Delta(x,\tau,z)$ is a matrix Hubbard-Stratonovich field.
As usual, at a mean-field level $\Delta_{ab}^0=U\langle 
\wpsi_b\psi_a-\delta_{ab}/2\rangle$, which is easily
solved to give $\Delta_{ab}^0(x)=\Delta_0(-1)^x\Lambda_{ab}$ with
$\Lambda_{ab}=S_p\delta_{ab}$. The massless fluctuations beyond the
mean-field theory can be represented by slowly-varying unitary rotations
of the ``staggered magnetization'' $\Delta_{ab}^0$. Ignoring the
massive modes associated with the amplitude fluctuations, we write
$$
\Delta_{ab}(x,\tau,z)= u_{ac}(x,\tau,z)\Delta^0_{cd}(x)
u^\dagger_{db}(x,\tau,z),
\ \  u\in SU(2n).
$$
In terms of the left ($\psi_L$) and right ($\psi_R$)
moving fermion fields defined in the continuum limit around the Fermi points 
in the strongly coupled $x$-direction, the action in Eq.~(\ref{s0}) can be
written as
\eqa
S_0&=&
\Tr(\wpsi_R\dminus\psi_R+\wpsi_L\dplus\psi_L)
-2i\delta t \Tr(\wpsi_R\psi_L-\wpsi_L\psi_R),
\nonumber \\
S_I&=&\Tr(\Delta_0^2/2U)-\Delta_0\Tr
(\wpsi_R u\Lambda u^\dagger\psi_L+\wpsi_L u\Lambda u^\dagger\psi_R),
\nonumber \\
S_\perp&=&i\tp\Tr\left[\wpsi_R(z)
\psi_L(z+1)+\wpsi_L(z)\psi_R(z+1)-{\rm c.c.}\right].
\nonumber
\eea
Here $\Tr$ stands for the trace over space-time as well as the replica and 
energy indices, $\partial_\pm=\dt\pm iv_F\dx$ with $v_F$ the Fermi velocity.
Next, we perform a local gauge transformation,
$
\psi_{L,R}^\prime=u^\dagger \psi_{L,R},
$
and define the pure SU(2n) gauge fields
$
A_{\pm}\equiv -iu^\dagger\partial_\pm u=A_\tau\pm iv_F A_x.
$
The action becomes (dropping the primes),
\eqa
S&=& \Tr\left[\wpsi_R(\dminus+iA_-)\psi_R+\wpsi_L(\dplus+iA_+)\psi_L\right]
\nonumber \\
&+&i\Delta^\prime\Tr\left[\wpsi_R e^{i\Lambda(\pi/2+
2\dtheta)}\psi_L-\wpsi_Le^{-i\Lambda(\pi/2+2\dtheta)}\psi_R\right]
\nonumber \\
&+&\Tr(\Delta_0^2/2U)+S_\perp(\psi_{L,R}\to u^\dagger\psi_{L,R}).
\label{s2}
\eea
where $\Delta^\prime=(\Delta_0^2+4\delta^2t^2)^{1/2}$ and
$2\dtheta={\rm tan}^{-1}(2\delta t/\Delta_0)$.

The final step is to integrate out the fermion fields.
In order to do so,
we need to bring the term proportional to $\Delta^\prime$ in Eq.~(\ref{s2})
to the standard mass term for Dirac fermions 
$i\Delta^\prime\Tr\left[\wpsi_R\psi_L-\wpsi_L\psi_R\right]$.
This can be done by the following chiral 
gauge transformation,
\eq
\psi_{R,(L),a}\to e^{(-)i\Lambda_{aa}(\pi/4+\dtheta)}\psi_{R,(L),a}.
\label{chiralrot}
\ee
As a result, we encounter the well-known
chiral anomaly \cite{fujikawa},
which arises from the
Jacobian associated with the transformation (\ref{chiralrot}) 
and leads to,
\eq
S_{\rm chiral}(A)={i\over\pi}\left({\pi\over4}+\dtheta\right)\Tr\uv 
\Lambda\dmu A_\nu,
\label{sc1}
\ee
where $\mu,\nu=x,\tau$, in the transformed action.
Now it is straightforward to integrate out the massive fermions
and obtain the effective action in terms of the gauge field. Using the
equalities $\Tr\uv\Lambda\dmu A_\nu=(i/4)\Tr\uv Q\dmu Q\dnu Q$
and $\Tr[A_\mu,\Lambda]^2=\Tr\dmu Q\dmu Q$, where
$Q\equiv u\Lambda u^\dagger\in U(2n)/U(n)\times U(n)$, we obtain,
\eq
S_{\rm eff}={\sxxo\over8}\Tr\dmu Q\dmu Q+{\sxyo\over8}\Tr\uv
Q\dmu Q\dnu Q
-{\szzo\over4\lambda^2}\Tr Q(z+1)Q(z)-h\Tr\Lambda Q.
\label{sq}
\ee
Here we have rescaled the coordinates by $\lambda x\to x$, $\lambda v_F\tau
\to\tau$ and $z\to z$. The coupling constants $\sigma_{\alpha,\beta}^0$
have the meaning of conductivities defined on the length scale cutoff
$\lambda$. For the present network model,
$\sxxo=(\Delta_0/\sqrt{\pi}\Delta^\prime)
^2$, $\sxyo=1/2+2\Delta\theta/\pi$, and
$\szzo=(\tp\Delta_0/\sqrt{\pi}v_F\Delta^\prime)^2$ 
for small $\tp/v_F$ and $\delta$, and
$h=\eta\Delta_0/v_FU\lambda^2$. The coupled-layers in the thermodynamic
limit is thus described by the zero-temperature properties of
the above (2+1)D quantum \sm.

For $\tp=0$, $\szzo=0$. Eq.~(\ref{sq}) reduces to $N$ independent
2D \sm s discovered by Pruisken and coworkers for the single layer
IQHE \cite{pruisken}. In this case, the term that couples to
$\sxyo$ becomes a topological quantity which 
produces the critical fixed points
at $(\sxx,\sxy)=({\rm const}, i+1/2)$ for the plateau transitions,
and the stable fixed points at $(\sxx,\sxy)=(0, i)$ for the
quantum Hall states in units of $e^2/h$. 
For the network model, the $i=0 \to1$
transition happens at $\delta=0$ where $\sxyo=1/2$. For $\sxyo\ne1/2$,
the conductances $(\sxx,\sxy)$ flow to $(0, 1)$ for $\delta>0$
and $(0,0)$ for $\delta<0$. At the transition, the dissipative
conductance has a critical value $\sxxc\simeq(0.58\pm.05)$
\cite{wbl,review}. Notice however, for finite $\szzo\ll\sxxo$, the system
is highly anisotropic but three-dimensional. 
The $\sxyo$-term, having two derivatives, no longer contains 
nontrivial topological 
contributions from slowly-varying field configurations on the scale of
the interlayer lattice spacing.

We now present a RG study of Eq.~(\ref{sq}). For weak interlayer tunnelings,
we follow the dimensional crossover analysis 
used in the O(3) \sm \  description of weakly coupled quantum spin 
chains \cite{affleck}.
The basic idea is that, since $R\equiv\szzo/\sxxo\ll1$, it is possible to
consider the renormalization of the coupling constants in Eq.~(\ref{sq})
in the 2D sector ($x,\tau$) independently until the renormalized couplings 
become comparable in all directions at 
a larger length scale $\lp$. The 3D isotropic RG is switched on 
beyond $\lp$. Since the scaling dimension of the $Q$-field is zero in
the replica limit, this crossover takes place when
$R\sxxo/\lambda^2\approx\sxxtd(\lp)/\lp^2$. 
One then takes the continuum limit in the $z$-direction by absorbing
the cutoff $\lp^{-2}$ into defining the derivatives and obtains
an isotropic 3D \sm \ action (plus the symmetry breaking term),
\eq
S_{\rm eff}^\prime={\sxx(\lp)\over8}\Tr\drho Q\drho Q
+{\sxy(\lp)\over8}\Tr\uv Q\dmu Q\dnu Q,
\label{sqi}
\ee
with $\rho=x,\tau,z$ and $\sigma_{\alpha\beta}(\lp)
=\sigma_{\alpha\beta}^{2d}(\lp)/\lp$, the conductivities at cutoff $\lp$.
The important point is that the latter are the bare coupling constants for the 
subsequent 3D RG \cite{affleck}.
Whether the system is in the insulating or metallic phase
is thus determined by the renormalized conductances 
at the end of the 2D RG.

Since the $\sxy$-term in the continuum action Eq.~(\ref{sqi}) is no 
longer topological, we performed perturbative RG calculations
to two-loop order to determine the flow of the conductance parameters.
This approach is valid in the metallic phase where
the bare conductivity is large for a large number of layers.
For general $n$, we found the recursion
relations for the conductivities under RG scale transformation,
$\lp\to b\lp$,
\eqa
\sxxp&=&\sxx\left[1-2n{1\over\sxx}I_d
+2(n^2+1){\epsilon\over d}{1\over\sigma_{xx}^2}I_d^2\right],
\label{sxxrg} \\
\sxyp&=&\sxy\left[1-2n{\epsilon\over d}
{1\over\sxx}I_d-8n^2{\epsilon\over d^2}{1\over\sigma_{xx}^2}I_d^2\right],
\label{sxyrg}
\eea
where $I_d=\int_{1/b\lp}^{1/\lp}d^dp/(2\pi)^d[1/(p^2+h)]$, $\epsilon=d-2$.
Notice that the corrections to $\sxyp$ vanish in
the replica limit $n\to0$, {\it i.e.}
{\em the Hall conductivity is unrenormalized 
in the 3D regime}. 
We will show later that this property is crucial for the quantization of
the Hall conductance.
This result should be contrasted to the one obtained in weak magnetic fields
where the Hall conductivity is found to renormalize
in the same way as the dissipative conductivity \cite{symp}.
Defining the dimensionless conductances $g_{\mu\nu}=\suvp\lp^\epsilon
b^\epsilon$ and $b=e^l$, Eq.~(\ref{sxxrg}) leads to the 
RG equation in the limit $n\to0$,
$dg_{xx}/dl=\epsilon g_{xx}-(4/dK_d^2)g_{xx}^{-1}$, $K_d=2^{d-1}\pi^{d/2}
\Gamma(d/2)$, consistent with the known result for the unitary \sm \  without
the $\sxy$-term in $2+\epsilon$ expansions \cite{brezin}.
For $d=3$, there is a nontrivial critical fixed point at
$g_c=\sqrt{4/dK_d^2\epsilon}=1/\sqrt{3}\pi^2$. It separates
a metallic phase with $g_{xx}(\lp)/g_c>1$ from an insulating phase,
where $g_{xx}(\lp)/g_c<1$. The dissipative conductivity vanishes at the
metal-insulator transition whereas the Hall conductivity remains 
close to its bare value at the beginning of the 3D RG.
The Hall conductance follows the simple Ohm's law, $dg_{xy}/dl=(d-2)g_{xy}$.

We now discuss the phase structure assuming the carrier density in each 
layer to be nominally the same. 
(i) For $\delta=0$, $\sxyo=1/2$. The individual layers are at the
critical point for the 2D plateau transition. During the initial
2D RG, $\sxytd$ does not renormalize and $\sxxtd$ flows
towards its finite critical value which is of order one.
Thus the crossover length $\lp\approx \lambda/\sqrt{R}$.
For small interlayer tunneling $R\ll1$, $\lp\gg\lambda$ such that
$\sxxtd(\lp)$ flows towards the fixed point value $\sxxtd(\infty)
=\sxxc\simeq0.55$ \cite{wbl,review}. The latter is greater
than the 3D critical conductance $g_c$ derived above. Thus we conclude
that when the Fermi energy is located at the critical point for
the 2D plateau transition, an arbitrarily small interlayer tunneling
leads to a 3D metallic state. 
A unique feature of the metallic
phase is that the Hall conductivity is unrenormalized and
remains close to $\sxy(\lp)$ down to low temperatures.
(ii) For $\delta\ne0$, the 2D RG scales towards the insulator/quantum Hall
fixed points, {\it i.e.} $\sxxtd\to0$, whereas $\sxytd\to0$ ($\delta<0$)
and $1$ ($\delta>0$) around the first plateau transition. 
The metallic phase is stable so long as 
$\sxxtd(\lp)> g_c$. Clearly, with decreasing (increasing)
$\tp$ ($\vert\delta\vert$), at a critical
$\tpc$ ($\delta_c$) where $\sxxtd(\lp)=g_c$, 
a metal-insulator transition takes place.
For $\tp<\tp^c$ (or $\vert\delta\vert>\delta_c$), the system is in the
3D insulating phase.
To determine the phase boundary,
notice that for $\delta\neq0$, a finite localization length develops 
in the 2D sector, $\xi_{2d}\propto\vert\delta\vert^{-\nu_{2d}}$ with 
$\nu_{2d}\simeq7/3$ \cite{cc,leewanghubbard,review}. Thus the 2D RG flow
stops at $\xi_{2d}$ beyond which a gap would develop in the 2D sector.
Setting the crossover length $\lp=\xi_{2d}$, one finds that the critical 
anisotropy $R_c\sim(\lambda/\xi_{2d})^2$, leading to the
phase boundary $\tpc\propto t\vert\delta\vert^{\nu_{2d}}$.
The width of the metallic phase
is then given by $W_\delta\propto(\tp/t)^{1/\nu_{2d}}$, consistent with
the numerical results of Chalker and Dohmen \cite{cd}.
(iii) From the above discussion, it is clear that
the metal-insulator transition is in the 3D unitary universality class
of the Anderson transition since the Hall conductivity appears to be
a 3D RG invariant. The two-loop RG equations imply
that the 3D localization length
diverges as $\xi_{3d}\propto \vert g_{xx}-g_c\vert^{-\nu_{3d}}$,
$\nu_{3d}=1/2\epsilon$. Simulations of various unitary models
give $\nu_{3d}=1.35\pm.15$ \cite{ohtsuki} which indeed agrees with
the numerical value $1.45\pm.25$ obtained directly from the layered 
network model \cite{cd}.

We next discuss the quantization of the Hall conductance in the 
insulating phases. In this case, during the first stage of the RG,
the Hall conductance $\sigma_{xy}^{2d}$ in Eq.~(\ref{sqi}) 
flows towards the 2D quantized values, {\it i.e.} 
$\sigma_{xy}^{2d}(\lambda^\prime)\to ie^2/h$,
for large anisotropy ($R\ll1$) such that $\lambda^\prime\to\infty$.
The quantization of the 3D Hall conductance is then
possible provided that $\sxy$ in Eq.~(\ref{sqi}) does not renormalize
in the 3D regime.
Restoring the discrete sum in the $z$-direction, 
{\it i.e.}, $(1/\lp)\int dz \to \sum_z$, this term becomes
$S_{xy}=N\sxytd(\lp)\int\!dxd\tau\tr\uv Q\dmu Q\dnu Q$,
leading to the Hall conductance quantization $\sxy\to iNe^2/h$.
The 3D quantum Hall states are therefore characterized by 
the resistance behaviors $\rho_{xx},\rho_{yy}\to0$, 
$\rho_{zz}\to\infty$, and $\rho_{xy}=(iN)^{-1}h/e^2$.
Due to the intervening metallic phases, the transitions between 
the quantized Hall plateaus comprise an insulator-metal and a subsequent 
metal-insulator transition, and have a finite width at low-temperatures.
These results are consistent with the experimental observations of the
IQHE in the 30-layer \cite{stormer} and the more recent 200-layer 
GaAs/AlGaAs structures \cite{brookswang}.

In contrast to 2D, where the electronic states are localized
at all energies except a critical set of zero measure,
it is the existence of metal-insulator transitions and 
the absence of localization corrections to the Hall conductivity that 
give rise to the quantization of the Hall conductance
in weakly coupled layered systems. This is supported by our two-loop 
RG results in Eqs~(\ref{sxxrg}) and (\ref{sxyrg}).
Although these results do not form a proof
to all orders in the perturbative RG, we believe that the evidence is 
sufficiently strong.

We now briefly discuss the possible topological effects not included in
the present analysis. The discrete (in the $z$-direction) action
in Eq.~(\ref{sq}) allows contributions from 
topologically-stable, point-singular field configurations, {\it i.e.} 
the hedgehogs at which the instanton number changes abruptly 
from one layer to the next \cite{read}. These contributions could in principle 
enter during the first stage of the RG and modify the bare parameters
of the continuum action in Eq.~(\ref{sqi}). The precise
effects of the hedgehogs in the replica limit is not understood
at the present time. Nevertheless we do not expect them to change 
the main results discussed above.

Finally, we consider the surface states in the quantum Hall phases
where the bulk localization length is very short. The edge state
supported by each layer couples together and forms an interesting
2D surface system decoupled from the bulk \cite{cd,surface}.
In the presence of boundaries, in addition to the bulk $\sxy$-term,
the action in Eq.~(\ref{sc1}) leads to an additional contribution
$(i\sxy/2)\oint d r_\mu \tr\Lambda A_\mu$,
where the integral is over the boundary of the sample at
$(x=0,L)$ while keeping the periodic boundary condition
in $\tau$. It is easy to show that this surface term can be written 
in terms of $Q(u,\tau,z)$, a smooth homotopy between
$Q(u=0)=Q(x=0)$ and $Q(u=1)=Q(x=L)$. The action on the surface is then,
\eq
S_{\rm sf}={\sxy\over4}\!\int_0^1\!du\Tr Q\partial_u Q
\dt Q +{\szzo\over8}\!\Tr\partial_z Q\partial_z Q.
\label{sf}
\ee
This action is identical to the coherent state path integral action
of an SU(2n) ferromagnetic Heisenberg spin chain with spin $S=\sxy/2$ 
and exchange $J=-\szzo/\sigma_{xy}^2$. The first term
in Eq.~(\ref{sf}) corresponds to the Berry phase term.
The spin quantization in this case results from the Hall conductance
quantization in the quantum Hall state.
The equations of motion of $S_{\rm sf}$ give the exactly known
one-magnon dispersion valid for all $n$, {\it i.e.}
$iq_\tau=\vert J\vert S q_z^2$.
By the analogy between the spin-spin correlation function
and the edge electron two-particle Green's function \cite{leewanghubbard}, 
it can be shown that this mode corresponds to the anisotropic 
diffusive mode on the surface, $i\omega\rho=-iq_\tau+(\szzo/2\sxy) q_z^2$.
Following Wegner \cite{wegner}, the latter
leads to the conductivities on the surface:
$\sigma_{zz}^{\rm sf}=\szzo/2\sxy$ and $\sigma_{\tau\tau}^{\rm sf}(\omega)
\propto i/\omega$. The present approach to the chiral surface state
is complimentary to those formulated using supersymmetric fields \cite{susy}.

The author thanks J.~S. Brooks, V. Dobrosavljevic, F. Gaitan, D-H Lee, and 
N. Read for many useful discussions, and Aspen Center for Physics for 
hospitality. This work was supported in part by an award from Research 
Corporation and the IHRP (500/5011) at the National High Magnetic Field 
Laboratory.
\vspace*{\fill}{

}

\begin{references}
\bibitem{qhe} For reviews, see {\it e.g.} {\sl The Quantum Hall
Effect}, edited by R.~E. Prange and S.~M. Girvin
(Springer-Verlag, 1990).
\bibitem{bh}B.~I. Halperin,
in {\sl Physical Phenomena at High Magnetic
Fields}, edited by E. Manousakis, {\it et al.} (Addison-Wesley, 1992).
\bibitem{stormer}H.~L. St\"omer, {\it et al.},
Phys. Rev. Lett. {\bf 56}, 85 (1986).
\bibitem{brookswang} J.~S. Brooks, {\it et. al.}, to be published.
\bibitem{organics} J.~S. Cooper, {\it et al.},
Phys. Rev. Lett. {\bf63}, 1984 (1989); S.~T. Hannahs, {\it et al.},
{\it ibid.} {\bf63}, 1988 (1989); S. Valfells, {\it et al.},
Phys. Rev. B{\bf54}, 16413 (1996).
\bibitem{leewanginter}
D-H Lee and Z. Wang, Phys. Rev. Lett. {\bf76}, 4014 (1996).
\bibitem{cc} 
J.~T. Chalker and P.~D. Coddington, J. Phys. C {\bf21},  2665 (1988);  
D-H Lee, Z. Wang and S.~A. Kivelson, Phys. Rev. Lett. {\bf70}, 4130 (1993). 
\bibitem{cd}J.~T. Chalker and A. Dohmen, Phys. Rev. Lett.
{\bf75}, 4496 (1995).
\bibitem{lee} D-H Lee, Phys. Rev. B{\bf 50}, 10788 (1994).
\bibitem{leewanghubbard}
D-H Lee and Z. Wang, Phil. Mag. Lett. {\bf73}, 145 (1996); and to be
published.
\bibitem{km} J. Kondev and J.~B. Marston, Report No. cond-mat/9612223.
\bibitem{fujikawa} K. Fujikawa, Phys. Rev. D{\bf21}, 2848 (1980).
\bibitem{pruisken} 
A.~M.~M. Pruisken, Chap. 5 in Ref.\cite{qhe}.
\bibitem{wbl} Z. Wang, B. Jovanovi\'c, and D-H Lee,  
Phys. Rev. Lett. {\bf77}, 4426 (1996).
\bibitem{review} B. Huckestein, Rev. Mod. Phys., {\bf67}, 357 (1995)
and references therein.
\bibitem{affleck} 
I. Affleck, M.~P. Gelfand and R.~R.~P. Singh,
J. Phys. A{\bf27}, 7313 (1994); I. Affleck and B.~I. Halperin, 
{\it ibid.} {\bf 29}, 2627 (1996);
Z. Wang, Phys. Rev. Lett. {\bf 78}, 126 (1997).
\bibitem{symp} M. Biafore, C. Castellani, and G. Kotliar,
Nucl. Phys. B{\bf340}, 617 (1990). X-F Wang, Z. Wang,
G. Kotliar, and C. Castellani, Phys. Rev. Lett. {\bf68}, 2504 (1992);
X-F Wang, Z. Wang, C. Castellani, M. Fabrizio,
and G. Kotliar, Nucl. Phys. B{\bf415}, 589 (1994).
\bibitem{brezin} E. Br\'ezin, S. Hikami, and J. Zinn-Justin,
Nucl. Phys. B{\bf 165}, 528 (1980).
\bibitem{ohtsuki} T. Ohtsuki, B. Kramer, and Y. Ono, J. Phys. Soc. Jpn.
{\bf62}, 224 (1993); M. Hennecke, B. Kramer, and T. Ohtuski,
Europhys. Lett. {\bf27}, 389 (1994).
\bibitem{read} S. Xiong, N. Read, and A.~D. Stone,
Phys. Rev. B{\bf56}, 3982 (1997).
\bibitem{surface} L. Balents and M.~P.~A. Fisher, 
Phys. Rev. Lett. {\bf76}, 2782 (1996);
Y. B. Kim, Phys. Rev. B{\bf53}, 16420 (1996).
\bibitem{wegner} F. Wegner, Phys. Rev. B{\bf19}, 783 (1979).
\bibitem{susy} L. Balents, M.~P.~A. Fisher, and M.~R.
Zirnbauer, Nucl. Phys. B{\bf483}, 601 (1996); I.~A. Gruzberg,
N. Read, and S. Sachdev, Phys. Rev. B{\bf55}, 10593 (1997).
\end{references}
\end{document}